\documentclass[12pt]{article}
\usepackage{amsmath}
\usepackage{amssymb}
\usepackage{graphicx}
\numberwithin{equation}{section} \textwidth=160mm
\begin{document}
\setcounter{page}{0} \thispagestyle{empty}
\begin{flushright}
\end{flushright}
\vspace*{2.0cm}
\begin{center}
{\large\bf Fermion Scattering off a CP-violating Bubble Wall \\
           in the Background of a Uniform Magnetic Field}
\end{center}
\vspace*{1cm}
\renewcommand{\thefootnote}{\fnsymbol{footnote}}
\begin{center}
{L. Campanelli$^{1,2}$\protect\footnote{Electronic address:
                {\tt campanelli@fe.infn.it}},
G.~L. Fogli$^{3,4}$\protect\footnote{Electronic address:
                {\tt Fogli@ba.infn.it}} and
L. Tedesco$^{3,4}$\protect\footnote{Electronic address:
                {\tt luigi.tedesco@ba.infn.it}}\\[0.5cm]
$^1${\em Dipartimento di Fisica,
         Universit\`a di Ferrara, I-44100 Ferrara, Italy}\\[0.2cm]
$^2${\em INFN - Sezione di Ferrara, I-44100 Ferrara, Italy}\\[0.2cm]
$^3${\em Dipartimento di Fisica,
         Universit\`a di Bari, I-70126 Bari, Italy}\\[0.2cm]
$^4${\em INFN - Sezione di Bari, I-70126 Bari, Italy}}
\end{center}
\vspace*{1.0cm}
\begin{center}
{September, 2004}
\end{center}
\vspace*{1.0cm}
\renewcommand{\abstractname}{\normalsize Abstract}
%
%
\begin{abstract}
In the scenario of the electroweak baryogenesis we consider the
dynamics of fermions with a spatially varying mass in presence of
a CP-violating bubble wall and a uniform magnetic field
perpendicular to the wall. The relevant quantity for baryogenesis,
$R_{R \rightarrow L} - R_{L \rightarrow R} \,$, is studied ($R_{R
\rightarrow L}$ and $R_{L \rightarrow R}$ being the reflection
coefficients for right-handed and left-handed chiral fermions,
respectively).
\end{abstract}
%
%
\vspace*{1.0cm}
\begin{flushleft} PACS number(s): 12.15.Ji, 11.30.Fs, 98.80.Cq
\end{flushleft}
\vfill
\newpage
%
%
\renewcommand{\thesection}{\normalsize{\arabic{section}.}}
\section{\normalsize{Introduction}}
\renewcommand{\thesection}{\arabic{section}}
The connection between particle physics and cosmology has received
much attention in the last years. In fact, cosmology can use the
predictions of particle physics in order to solve cosmological
problems. The origin of the asymmetry between matter-antimatter is
one of those important open questions in cosmology. As Sakharov
pointed out more than thirty years ago \cite{sakharov}, it is
possible to start with a Universe in a baryon symmetry state and
later dynamically to generate a net asymmetry by particle
interactions. He postulated three conditions in order to explain
the observed baryon excess: 1) {\it Baryon number
non-conservation}; this violation occurs in the Standard Model
\cite{hooft}-\cite{kuzmin} through the axial anomaly, but the rate
of baryon number non-conserving processes are exponentially
suppressed at zero temperature. 2) {\it C and CP-violation}; this
violation has been observed in the neutral K meson system and
recently in the B meson decay \cite{aubert}. 3) {\it Departure
from thermal equilibrium}; in fact, in thermal equilibrium
particles and antiparticles will have the same number density.
\\
The Standard Model has in principle all the necessary ingredients
for Sakharov conditions for the generation of the baryon asymmetry
of the Universe \cite{kuzmin}. In particular the electroweak
baryogenesis is a good candidate because give us an explanation in
terms of experimentally accessible physics. Essentially the baryon
asymmetry problem is twofold to understand: 1) why the observable
density of baryons in the Universe is much greater than the
density of the antibaryons and 2) why the density of baryons is
much less than density of photons. Indeed, the measurement of the
ratio of baryon density $n_B$ (defined as the number density of
baryons $n_b$ minus the number density of antibaryons
$n_{\bar{b}}$) and the photon density $n_\gamma$, at the present
time, recently has been determined from the measurement of the
relative heights of the first two CMB acoustic peaks by WMAP
satellite. It takes the value \cite{bennet}
\begin{equation}
\label{eq1.1} 5.9 \times 10^{-10} \leq \, \frac{n_B}{n_{\gamma}}
\, \leq 6.4 \times 10^{-10}.
\end{equation}
CP-violation is crucial to generate the baryon asymmetry. It is
well known, however, that the amount of CP-violation in the quark
sector of the standard model cannot account for the observed
baryon asymmetry \cite{{shap},{farrar}}. In fact, the strength of
CP-violation coming from the CKM phase is inadequate by 10-12
orders of magnitude ($n_B/n_\gamma \sim 10^{-20})$ \cite{barr}.
\\
Another question is connected with the third Sakharov condition.
The baryogenesis necessarily requires non-equilibrium physics. In
order to avoid the so called washout after the phase transition,
it is necessary to have $v(T_c)/T_c \geq 1$, where $v(T_c)$ is the
vacuum expectation value of the broken Higgs field. This condition
ensures the surviving of the baryon asymmetry after the walls has
passed and indicates that the electroweak phase transition should
be, if strong enough, of the first order \cite{carrington}.
Moreover, lattice studies have shown that for any Higgs mass, the
phase transition would be so weak that sphaleron interactions are
in equilibrium in the broken phase. This implies that the baryon
asymmetry goes to zero immediately after its generation
\cite{kajantie}. On the other hand, if we use one loop high
temperature effective potential we have an Higgs boson mass $M_H
\leq 60 \, \text{Gev}$ \cite{shap2} which is in contrast with the
experimental lower bound reached $M_H > 114,4 \, \text{Gev}$ \cite
{hagiwara}. Hence, the magnitude of CP-violation and the
experimental limits on Higgs mass contradict the possibility to
have an asymmetry of the experimentally observed order within the
framework of the Minimal Standard Model and one is lead to
consider extensions of the Standard Model. If we demands that the
electroweak phase transition has to be of first order, we have a
mechanism for the transition called bubble nucleation. It proceeds
by formation and expansion of bubbles of new phase within the old
ones. The study of the propagation of the bubble has been object
of much attention as from the early works \cite{langer},
\cite{linde}. The first order provided a natural way to depart
from equilibrium that takes places near the wall of the expanding
bubbles. Moreover, in the unbroken phase where the sphaleron is
active, the asymmetries of some local charges are converted into a
baryon asymmetry. The baryon number flows into the new phase and
the baryogenesis is produced. In this scenario of electroweak
baryogenesis there are two interesting mechanisms that involve
interacting fermions with CP-violating background of the bubble.
In both cases, it is possible to have a net baryon flux by using
fermion scattering off bubble walls \cite{cohen2}: {\it The local
baryogenesis} \cite{mclerran}, \cite{turok} in which the
CP-violating and B-violating processes are near or at the bubble
wall. 2) {\it Non local baryogenesis} \cite{cohen2} (or charge
transport mechanism) in which CP-violation and baryon number
violation are separated from one another.
\\
In our work we consider this last kind of electroweak
baryogenesis. Moreover, the non local baryogenesis is effective
for thin-wall regime \cite{cohen2}-\cite{joyce}. Hence, it is
possible to treat as free the fermions that may only interact in a
small region with Higgs background field. As a consequence of
CP-violation on the wall, we will have asymmetric reflection and
transmission, and the particles injected into the broken phase
diffuse in the bubble. The correct way to treat the generation of
the asymmetry in front of the bubble is controversial. A very
simple mode is to consider an extra source of CP-violation,
parameterized assigning a spatially varying complex mass to the
Higgs field:
\begin{equation}
\label{eq1.1} m(t,\vec x) = m_R(t,\vec x) + i m_I(t,\vec x).
\end{equation}
The imaginary part $m_I$ gives rise to CP-violation. The problem
to solve Dirac equation in the background of the bubble wall
without CP-violation \cite{ayala} and with CP-violation
\cite{funakubo} has been already studied. The authors also have
studied the case of fermion scattering in the background of the
bubble wall without CP-violation and in presence of a constant
magnetic field perpendicular to the wall \cite{cea}.
\\
It is important to stress that the introduction of a magnetic
field is not an academic exercise because astronomical
observations show the presence of a cosmic magnetic field in all
galaxies \cite{dolgov}. There are some possible mechanism for
generating a cosmological magnetic field in the early Universe.
One of these is related to the first order electroweak phase
transition \cite{hogan}, in which, bubbles of the new phase
expanding in the old ones generate electric currents that, in
turns, produce magnetic fields \cite{vachaspati}.
\\
In this paper we study and solve the Dirac equation in presence of
a CP-violating planar bubble wall with a constant magnetic field
perpendicular to the wall. We consider the scattering
perpendicular to the wall and the radius of the bubble very large.
Neglecting the time dependence of mass terms, we shall work in the
hypothesis that $m(t,\vec x)$ only depends by the $z$-coordinate
perpendicular to the wall: $m(t,\vec x)= m(z)$. The real part goes
to zero when $z \rightarrow - \infty$ (symmetric phase) and goes
to $m_0$ (broken phase) when $z \rightarrow + \infty$, where $m_0$
is fermion mass. We calculate the reflection coefficients, $R_{R
\rightarrow L}$ and $R_{L \rightarrow R}$, of left-handed and
right-handed fermions, respectively, at a radially expanding
bubble. The difference $\Delta R = R_{R \rightarrow L} - R_{L
\rightarrow R}$ is important as regards the baryon asymmetry
\cite{Nelson} in cosmology.
%
%
\renewcommand{\thesection}{\normalsize{\arabic{section}.}}
\section{\normalsize{CP-violating Dirac equation}}
\renewcommand{\thesection}{\arabic{section}}
In this Section we study the scattering of Dirac fermions off
CP-violating bubble walls in the presence of an electromagnetic
field $A_\mu$. Thus, assuming that a fermion $\Psi$ is coupled to
a complex scalar field $\Phi$ through a Yukawa interaction with
coupling $g_Y$, the Dirac equation reads:
\begin{equation}
\label{eq2.1} [\, i / \!\!\!\partial - m(t,\vec x) P_R -
m^{*}(t,\vec x) P_L - e \: / \!\!\!\!\!\:\!A \, ] \, \Psi(t,\vec
x) = 0,
\end{equation}
where $e$ is the electric charge,
$P_R = (1 + \gamma_5)/2$ and $P_L =(1-\gamma_5)/2$,
are the right-handed and left-handed projection operators such
that $P_R P_L = P_L P_R = 0$, $P_R^2 = P_R$, $P_L^2 = P_L$,
and
\begin{equation}
m(t,\vec x)= - g_Y \langle \Phi(t,\vec x) \rangle
\end{equation}
is a complex function of space-time. We only consider the motion
of fermions perpendicular to the wall, and the radius of the
bubble very large in order to approximate the surface by a flat
wall perpendicular to the $z$-axis. Hence, we shall consider
$m(t,\vec x) = m(z)$. In order to solve Eq.~(\ref{eq2.1}), we make
the following ansatz:
\begin{equation}
\label{eq2.2} \Psi(t,z) = [i / \!\!\!\partial + m(z) P_R +
m^{*}(z) P_L - e \: / \!\!\!\!\!\:\!A \, ] \, e^{-i \sigma E t} \,
\psi_E(z),
\end{equation}
where $\sigma = \pm 1$ for positive and negative energy solutions,
respectively. Moreover, we assume that $A_\mu$ corresponds to a
constant and uniform magnetic field directed along the $z$-axis
with strength $B$. We can then choose the Landau gauge
\begin{equation}
A_\mu = (0,0,-Bx,0).
\end{equation}
Inserting Eq.~(\ref{eq2.2}) into Eq.~(\ref{eq2.1}), we get
\begin{equation}
\label{eq2.3} \left[ E^2 + \frac{d^2}{dz^2} - |m|^2 + i \gamma^3
\frac{d m_R}{dz} + \gamma^3 \gamma_5 \frac{d m_I}{d z} + ieB
\gamma^1 \gamma^2 \right] \psi_E(z) = 0.
\end{equation}
It is useful to introduce a parameter mass term $a$, ($1/a$ being
the characteristic size of the thickness of the wall) and define
\begin{eqnarray}
\label{eq2.4} m_R(z) \!\!& = &\!\! m_0 f(az) = m_0 f(x),
\\
\label{eq2.5} m_I(z) \!\!& = &\!\! m_0 g(az) = m_0 g(x),
\end{eqnarray}
where
\begin{equation}
\label{eq2.6} x = az, \;\;\; \tau = a t, \;\;\;\; \epsilon = E/a,
\;\;\;\; \xi = m_0/a, \;\;\;\; b = eB/a^2.
\end{equation}
Taking into account Eqs.~(\ref{eq2.4})-(\ref{eq2.6}), the ansatz
(\ref{eq2.2}) and Eq.~(\ref{eq2.3}) become, respectively
\begin{eqnarray}
\label{eq2.8} & & \Psi(\tau,x) = \left[ \sigma \epsilon \gamma^0 +
i \gamma^3 \frac{d}{dx} + \xi f - i \xi g \gamma_5 + bx \gamma^2
\right] e^{-i \sigma \epsilon \tau} \psi_{\epsilon}(x),
\\
\label{eq2.9} & & \left[ \epsilon^2 + \frac {d^2} {dx^2} - \xi^2
(f^2 + g^2) +i\, \xi \gamma^3 f' - \xi \, \gamma_5 \gamma^3 g'  +
ib \, \gamma^1 \gamma^2 \right] \psi_{\epsilon}(x) = 0 ,
\end{eqnarray}
where a prime denotes differentiation with respect to $x$. Let us
suppose that $|g(x)| \ll 1$. The small value of $g$ allows to
consider it as a perturbation and to consider only first order
terms in $g$. We assume that
\begin{equation}
\label{eq2.10} \lim_{x \rightarrow + \infty} f(x)  = 1,  \;\;\;\;
\lim_{x \rightarrow - \infty} f(x)  = 0.
\end{equation}
In first order approximation we write
\begin{equation}
\label{eq2.11} \psi_{\epsilon}(x) = \psi^{(0)}(x)  +
\psi^{(1)}(x).
\end{equation}
The wave function $\psi^{(0)}$ is a solution of the unperturbed
equation
\begin{equation}
\label{eq2.12} \left[ \epsilon^2 + \frac{d^2}{dx^2} - \xi^2 f^2 +
i \xi \gamma^3 f' + ib \gamma^1 \gamma^2 \right] \psi^{(0)}(x) =
0,
\end{equation}
which is obtained from Eq.~(\ref{eq2.9}) with $g=0$. The
perturbation $\psi^{(1)}$ can be calculated as
\begin{equation}
\label{eq2.13} \psi^{(1)}(x) = \int \! dx' \, G(x,x') V(x') \,
\psi^{(0)} (x') \, ,
\end{equation}
where $V(x) = - \xi g(x') \gamma_5 \gamma^3$, and $G(x,x')$ is a
Green function that satisfies the following equation,
\begin{equation}
\label{eq2.14} \left[ \epsilon^2 + \frac{d^2}{dx^2} - \xi^2 f^2 +
i \xi f' \gamma^3 + ib \gamma^1 \gamma^2 \right]_{\alpha \beta}
\!\! G_{\beta \gamma} (x,x') = - \delta_{\alpha \gamma}
\delta(x-x').
\end{equation}
Inserting Eq.~(\ref{eq2.11}) in Eq.~(\ref{eq2.9}) we have
\begin{equation}
\label{eq2.15} \Psi(\tau,x) \simeq \left[ \left(\sigma \epsilon
\gamma^0 + i \gamma^3 \frac{d}{dx} + \xi f + bx \gamma^2  \right)
(\psi^{(0)} + \psi^{(1)}) + i \xi g \gamma_5 \psi^{(0)} \right] \!
e^{-i \sigma \epsilon \tau} ,
\end{equation}
to the first order in $g$. In order to obtain $\psi^{(0)}$ we have
to solve Eq.~(\ref{eq2.12}) (see Refs. \cite{funakubo},
\cite{cea}). Let us expand $\psi^{(0)}$ in terms of the
eigenstates of $\gamma^3$:
\begin{equation}
\label{eq2.16} \psi^{(0)}(x) = \phi_{\pm}^{(s)}(x) \, u_{\pm}^s \,
,
\end{equation}
where $s = 1,2$, and
\begin{equation}
\label{eq2.17} u_{\pm}^1 = \frac{1}{\sqrt{2}} \! \left(
\begin{array}{c}
1 \\ 0 \\ \pm i \\ 0
\end{array}
\right) \! , \;\;\;\; u_{\pm}^2 = \frac{1}{\sqrt{2}} \! \left(
\begin{array}{c}
0 \\ 1 \\ 0 \\ \mp i
\end{array}
\right) \! .
\end{equation}
The spinors $u_{\pm}^s$ satisfy the relations
\begin{eqnarray}
\label{eq2.19} \gamma^0 u^s_{\pm} \!\!& = &\!\! u^s_{\mp} \, ,
\\
\gamma^1 \gamma^2 u_{\pm}^s \!\!& = &\!\! i (-1)^s u_{\pm}^s \, ,
\\
\gamma^3 u_{\pm}^s \!\!& = &\!\! \pm i u_{\pm}^s \, ,
\\
\gamma_5 u^s_{\pm} \!\!& = &\!\! \mp i (-1)^s u^s_{\mp} \, ,
\end{eqnarray}
that will be useful in the following. Inserting Eq.~(\ref{eq2.16})
into Eq.~(\ref{eq2.12}), we get
\begin{equation}
\label{eq2.21} \left[\epsilon^2 + \frac{d^2}{dx^2} - \xi^2 f^2 \mp
\xi f' - (-1)^s b \right] \! \phi_{\pm}^{(s)}(x) = 0 .
\end{equation}
We indicate the independent solutions of Eq.~(\ref{eq2.21}) as
$\phi_{\pm}^{(+\alpha_s)}(x)$ and $\phi_{\pm}^{(-\alpha_s)}(x)$.
The asymptotic properties of $\phi_{\pm}^{(+\alpha_s)}$ and
$\phi_{\pm}^{(-\alpha_s)}$ are, respectively
\begin{eqnarray}
\label{eq2.22} \phi_{\pm}^{(+\alpha_s)}(x)  & \rightarrow &
           \begin{cases}
             e^{+\alpha_s x}
             \;\;\;\;\;\;\;\;\;\;\;\;\;\;\;\;\;\;\;\;\;\;\;\;
             \;\;\;\;\;\;\;\;\;\;\;\;\;\;\;\;\;\;\;\;\;\;\;\;
             \;\;\;\;\;\;
             \text{for} \; x \rightarrow + \infty ,
             \\
             \gamma_{\pm}(\alpha_s,\beta_s) \, e^{\beta_s x}+
             \gamma_{\pm}(\alpha_s,-\beta_s) \, e^{-\beta_s x}
             \;\;\;\;\;\,\;\;\;\;
             \text{for} \; x \rightarrow - \infty ,
           \end{cases}
\\ \nonumber
\\
\label{eq2.24} \phi_{\pm}^{(-\alpha_s)}(x) & \rightarrow &
           \begin{cases}
             e^{-\alpha_s x}
             \;\;\;\;\;\;\;\;\;\;\;\;\;\;\;\;\;\;\;\;\;\;\;\;
             \;\;\;\;\;\;\;\;\;\;\;\;\;\;\;\;\;\;\;\;\;\;\;\;
             \;\;\;\;\;\;
             \text{for} \; x \rightarrow + \infty ,
             \\
             \gamma_{\pm}(-\alpha_s,\beta_s) \, e^{\beta_s x}+
             \gamma_{\pm}(-\alpha_s,-\beta_s) \, e^{-\beta_s x} \;\;\;\;
             \text{for} \; x \rightarrow - \infty ,
\end{cases}
\end{eqnarray}
where
\begin{eqnarray}
\label{eq2.26} \alpha_s \!\!& = &\!\! i \sqrt{\epsilon^2 - \xi^2 -
(-1)^s b} \, ,
\\
\beta_s \!\!& = &\!\! i \sqrt{\epsilon^2 - (-1)^s \, b} \, ,
\end{eqnarray}
and $\gamma_{\pm}(\alpha_s,\beta_s)$ are constants such that
$\gamma_{\pm}(\alpha_s,\beta_s)^* =
\gamma_{\pm}(-\alpha_s,-\beta_s)$. The general unperturbed
solution is
\begin{equation}
\label{eq2.29} \psi^{(0)}(x) = \sum_{s = 1,2} \, [
A_+^{(-\alpha_s)} \phi^{(-\alpha_s)}_+ (x) + A_+^{(+\alpha_s)}
\phi^{(+\alpha_s)}_+ (x) ] \, .
\end{equation}
The incident wave function coming from $x = -\infty$ is reflected
and transmitted, while at $x = + \infty$ there is only the
transmitted wave. Thus, we have $A^{(-\alpha_s)} = 0$ for $\sigma
= +1$ and $A^{(+\alpha_s)} = 0$ for $\sigma = -1$.
\\
Following Ref.~\cite{funakubo}, in order to calculate the Green
function $G(x,x')$, we introduce a unitary matrix
\begin{equation}
\label{eq2.30} U = ( \, u^1_+  \;\, u^1_- \;\, u^2_+ \;\, u^2_-
\,) = \frac{1}{\sqrt{2}} \left(
\begin{array}{clcr}
1 &  \; 1  \!&  \; 0  & 0 \\
0 &  \; 0  \!&  \; 1  & 1 \\
i &    -i  \!&  \; 0  & 0 \\
0 &  \; 0  \!&    -i  & i
\end{array}
\right) \! ,
\end{equation}
and we write $G(x,x')$ as
\begin{equation}
\label{eq2.31} G(x,x') = U \left(
\begin{array}{clcr}
G_+(x,x') &   &  &  \\
 &  G_-(x,x') &  &  \\
 &  & G_+(x,x')  &  \\
 & & & G_-(x,x')
\end{array}
\right) U^{-1},
\end{equation}
where
\begin{equation}
\label{eq2.32} \left[ \frac{d^2}{d x^2} \mp \xi f' - \xi^2 f^2 +
\epsilon^2 - (-1)^s \, b \right] \! G_{\pm} (x,x') = - \delta
(x-x').
\end{equation}
Following the standard method \cite{hilbert}, we find for the
Green function the expression
\begin{equation}
\label{eq2.33} G_{\pm}^{(s , \sigma)}(x,x') =
                         \begin{cases}
                          - \frac{\sigma}{2 \alpha_s}
                          [\phi_{\pm}^{(-\sigma, \alpha_s)}(x) +
                          c_{\pm}^{(s, \sigma)}
                          \phi_{\pm}^{(+\sigma,
                          \alpha_s)}(x)] \, \phi_{\pm}^{(-\sigma,
                          \alpha_s)}(x') \;\;\;\;\, \mbox{if} \;\; x < x',
                          \\
                          \phantom{xxxxxx}
                          \\
                          - \frac{\sigma}{2 \alpha_s}
                          [\phi_{\pm}^{(-\sigma, \alpha_s)}(x') +
                          c_{\pm}^{(s, \sigma)}
                          \phi_{\pm}^{(+\sigma,
                          \alpha_s)}(x')] \, \phi_{\pm}^{(-\sigma,
                          \alpha_s)}(x) \;\;\;\; \mbox{if} \;\; x > x'.
                          \end{cases}
\end{equation}
Therefore, if we know the Green function and wave function
$\psi^{(0)}$ it is possible to find, by Eq.~(\ref{eq2.13}), the
wave function $\psi^{(1)}$:
\begin{eqnarray}
\label{eq2.34} \psi^{(1)}_s(x) \!\!& = &\!\! \, A^{(+)}_s
\frac{(-1)^s \, \xi}{2 \alpha_s} \, u^{(s)}_{-} \nonumber
\\
\!\!& \times &\!\! \left\{ \phi^{(+\alpha_s)}_{-} (x)
\int_{-\infty}^x \!\! dx' g'(x') \! \left[
\phi_-^{(-\alpha_s)}(x') + c^{(+)}_{-} \phi^{(+ \alpha_s)}_{-}(x')
\right] \! \phi_+^{(+\alpha_s)}(x') \right. \nonumber
\\
\!\!& + &\!\! \left. \left[ \phi_{-}^{(-\alpha_s)}(x) +
c^{(+)}_{-} \phi^{(+\alpha_s)}_{-} (x) \right] \int_x^{\infty}
\!\! dx' g'(x') \, \phi^{(+\alpha_s)}_{-} (x') \, \phi_+^{(+
\alpha_s)}(x') \right\} .
\end{eqnarray}
Let us introduce the following quantities:
\begin{eqnarray}
\label{eq2.35} I_1^{(s)} \!\!& = &\!\! \int_{-\infty}^{+\infty}
\!\! dx \, g'(x) \, \phi^{(-\alpha_s)}_- (x) \,
\phi^{(+\alpha_s)}_+(x) \, ,
\\
\label{eq2.35a} I_2^{(s)} \!\!& = &\!\! \int_{-\infty}^{+\infty}
\!\! dx \, g'(x) \, \phi^{(+\alpha_s)}_- (x) \,
\phi^{(+\alpha_s)}_+(x) \, .
\end{eqnarray}
In Appendix A we shall calculate the asymptotic expression of the
transmitted, incident and reflected wave functions (for
definiteness, let us consider the case $\sigma = +1$). They are
respectively,
\begin{eqnarray}
\label{eq2.36} [\Psi_s(\tau,x)]^{\mbox{{\scriptsize
tran}}}_{\sigma = +1} \!\!& = &\!\! A^{(+)}_s \, e^{-i \epsilon
\tau + \alpha_s x} \nonumber
\\
\!\!& \times &\!\! \left\{ \! (\xi - \alpha_s) \!\! \left[ 1 +
\frac{(-1)^s \xi (\xi + \alpha_s) I_1^{(s)}}{2 \epsilon \alpha_s}
+ \frac{\xi b \, ( I_1^{(s)} + c^{(+)}_{-} I_2^{(s)})}{2 \epsilon
\alpha_s (\xi - \alpha_s)} + \frac{(-1)^s \xi g_{-}}{2 \epsilon}
\right] \! u^s_{+} \right. \nonumber
\\
\!\!& + &\!\! \left. \epsilon \! \left[  1 +  \frac{(-1)^s \xi
(\xi + \alpha_s) I_1^{(s)}}{2 \epsilon \alpha_s} - \frac{(-1)^s
\xi g_{+}}{ \epsilon} + \frac{(-1)^s \xi g_{-}}{2 \epsilon}
\right] \! u^s_{-} \right\} ,
\\ \nonumber
\\
\label{eq2.37} {[\Psi_s(\tau,x)]}^{\mbox{{\scriptsize
inc}}}_{\sigma = +1} \!\!& = &\!\! A^{(+)}_s
\gamma_{+}(\alpha_s,\beta_s) \, e^{-i \epsilon \tau + \beta_s x}
\nonumber
\\
\!\!& \times &\!\! \left\{ - \beta_s \! \left[ 1 - \frac{(-1)^s
\xi \epsilon \, I_2^{(s)}}{2 \alpha_s \beta_s} \,
\frac{\gamma_{-}(-\alpha_s,\beta_s)}{\gamma_{+}(\alpha_s,\beta_s)}
+ \frac{\xi b \, c^{(+)}_{-} I_2^{(s)}}{2 \epsilon \alpha_s (\xi -
\alpha_s)} + \frac{(-1)^s \xi g_{-}}{2 \epsilon} \right] \!
u^s_{+} \right. \nonumber
\\
\!\!& + &\!\! \left. \epsilon \! \left[ 1 + \frac{(-1)^s \xi
\beta_s I_2^{(s)}}{2 \epsilon \alpha_s} \,
\frac{\gamma_{-}(-\alpha_s,\beta_s)}{\gamma_{+}(\alpha_s,\beta_s)}
- \frac{(-1)^s \xi g_{-}}{2 \epsilon} \right] \! u^s_{-} \right\}
,
\\ \nonumber
\\
\label{eq2.38} {[\Psi_s(\tau,x)]}^{\mbox{{\scriptsize
refl}}}_{\sigma = +1} \!\!& = &\!\! \left .
{[\Psi_s(\tau,x)]}^{\mbox{{\scriptsize inc}}}_{\sigma = +1}
\right|_{\beta_s \rightarrow - \beta_s},
\end{eqnarray}
where $g_{\pm} = \lim_{x \rightarrow \pm \infty} g(x)$.
From Eqs.~(\ref{eq2.36})-(\ref{eq2.38}) we calculate the vectorial
current
$j^{\mu}_V = {\bar{\Psi}} \gamma^{\mu} \Psi$ and the axial current
$j^{\mu}_A = \bar{\Psi} \gamma^{\mu} \gamma_5 \Psi$.
After some manipulations we obtain
\begin{eqnarray}
\label{eq2.40} (j^3_{V,s})^{\mbox{{\scriptsize tran}}} \!\!& =
&\!\! 2 \epsilon |A_s^{(+)}|^2 |\alpha_s| \, (1+
\delta^{\mbox{{\scriptsize tran}}}) \, ,
\\
(j^3_{V,s})^{\mbox{{\scriptsize inc}}} \!\!& = &\!\! 2 \epsilon
|A^{(+)}_s|^2 |\gamma_+(\alpha_s, \beta_s)|^2 |\beta_s| \, (1+
\delta^{\mbox{{\scriptsize inc}}}_{1}+ \delta^{\mbox{{\scriptsize
inc}}}_{2}) \, ,
\\
(j^3_{V,s})^{\mbox{{\scriptsize refl}}} \!\!& = &\!\! - \left.
(j^3_{V,s})^{\mbox{{\scriptsize inc}}} \right|_{\beta_s
\rightarrow - \beta_s} \! ,
\end{eqnarray}
where
\begin{eqnarray}
\label{eq2.43} \delta ^{\mbox{{\scriptsize tran}}} \!\!& = &\!\!
\frac {\xi b} {2 \epsilon |\alpha|^2_s} \, \mbox{Re} \! \left[
I_1^{(s)} + c^{(+)}_- I^{(s)}_2 \right] \! ,
\\
\delta^{\mbox{{\scriptsize inc}}}_{1} \!\!& = &\!\! (-1)^s \,
\frac{\xi |\beta_s|}{\epsilon |\alpha_s|} \, \mbox{Re} \! \left[
\frac {\gamma_-(- \alpha_s, \beta_s)} {\gamma_+
(\alpha_s,\beta_s)} \, I^{(s)}_2 \right] \! ,
\\
\delta^{\mbox{{\scriptsize inc}}}_{2} \!\!& = &\!\! \frac{\xi b}{2
\epsilon |\alpha_s| |\beta_s|} \left\{ \mbox{Re} \! \left[ \frac
{\gamma_-(- \alpha_s, \beta_s)} {\gamma_+(\alpha_s, \beta_s)} \,
I_2^{(s)} \right] + \mbox{Im} \! \left[ \frac{i \beta_s
c^{(+)}_-}{\alpha_s - \xi} \, I_2^{(s)} \right] \right\} \! .
\end{eqnarray}
The transmission and reflection coefficients are
\begin{eqnarray}
\label{eq2.46} T_{s,B} \!\!& = &\!\!
\frac{(j^3_{V,s})^{\mbox{{\scriptsize
tran}}}}{(j^3_{V,s})^{\mbox{{\scriptsize inc}}}} \, = \,
T^{(0)}_{s,B} \, ( 1 + \delta^{\mbox{{\scriptsize tran}}} -
\delta^{\mbox{{\scriptsize inc}}}_1 - \delta^{\mbox{{\scriptsize
inc}}}_2 \,) ,
\\
\label{eq2.47} R_{s,B} \!\!& = &\!\! -
\frac{(j^3_{V,s})^{\mbox{{\scriptsize
refl}}}}{(j^3_{V,s})^{\mbox{{\scriptsize inc}}}} \, = \,
R^{(0)}_{s,B} \, (1+ \delta^{\mbox{{\scriptsize refl}}}_1 +
\delta^{\mbox{{\scriptsize refl}}}_2 - \delta^{\mbox{{\scriptsize
inc}}}_1 - \delta^{\mbox{{\scriptsize inc}}}_2) ,
\end{eqnarray}
where
\begin{equation}
\label{eq2.48} T^{(0)}_{s,B} = \frac{|\alpha_s|}{|\beta_s| |\,
\gamma_+(\alpha_s, \beta_s)\, |^2} \, , \;\;\;\;\; R_{s,B}^{(0)}=
\left| \frac{\gamma_+(\alpha_s, -\beta_s)}{\gamma_+(\alpha_s,
\beta_s)} \right|^2 \!\! ,
\end{equation}
are the transmission and reflection coefficients in the absence of
CP-violation \cite{cea}, and
\begin{equation}
\label{eq2.50} \left. \delta^{\mbox{{\scriptsize refl}}}_1 = -
\delta^{\mbox{{\scriptsize inc}}}_1 \right|_{\beta_s \rightarrow -
\beta_s} \! , \;\;\;\;\; \left. \delta^{\mbox{{\scriptsize
refl}}}_2 = - \delta^{\mbox{{\scriptsize inc}}}_2 \right|_{\beta_s
\rightarrow - \beta_s} \! .
\end{equation}
The unitary condition, $R^{(0)}_{s,B} + T^{(0)}_{s,B} = 1$, holds
as it should be. We find the expression for $c^{(+)}_-$ by
requiring the unitary condition for the reflection and
transmission coefficients $R_{s,B}$ and $T_{s,B}$. Imposing that
\begin{equation}
R_{s,B} + T_{s,B} = 1,
\end{equation}
we get
\begin{equation}
\label{eq2.51} c^{(+)}_{-} = - \frac{1}{\xi} \,
\frac{\mbox{Re}[I_1^{(s)}]}{\mbox{Re} [I_2^{(s)} / (\xi -
\alpha_s)]} \: .
\end{equation}
Now, we calculate the axial currents
$(j^3_{A,s})^{\mbox{{\scriptsize inc}}}$ and
$(j^3_{A,s})^{\mbox{{\scriptsize refl}}}$, analogously to the
vectorial current. We get the following expressions:
\begin{eqnarray}
\label{eq2.52} (j^3_{A,s})^{\mbox{{\scriptsize inc}}} \!\!& =
&\!\! (-1)^{s+1} |A_s^{(+)}|^2 |\gamma_+(\alpha_s, \beta_s)|^2
\nonumber
\\
\!\!& \times &\!\! \left[ \epsilon^2 (1 +
\delta^{\mbox{{\scriptsize inc}}}_1 \,) + |\beta_s|^2 (1 +
\delta^{\mbox{{\scriptsize inc}}}_1 + 2 \delta^{\mbox{{\scriptsize
inc}}}_2 \,) \right] \! ,
\\ \nonumber
\\
\label{eq2.53} (j^3_{A,s})^{\mbox{{\scriptsize refl}}} \!\!& =
&\!\! (j^3_{A,s})^{\mbox{{\scriptsize inc}}} \left|_{\beta_s
\rightarrow - \beta_s} \right. \! ,
\end{eqnarray}
where, here and in the following (see Eqs.~(\ref{eq2.57}),
(\ref{eq2.58}) and (\ref{eq2.59})), we set $g_{-} = 0$.

Finally, we are able to calculate the relevant quantity for the
generation of a cosmological baryon asymmetry,
\begin{equation}
\label{eq2.54} \Delta R^{(s)} = R_{R \rightarrow L}^{(s)} - R_{L
\rightarrow R}^{(s)} \, ,
\end{equation}
where
\begin{eqnarray}
\label{eq2.55} R_{R \rightarrow L}^{(s)} \!\!& = &\!\! -
\frac{(j^3_{L,s})^{\mbox{{\scriptsize
refl}}}}{(j^3_{R,s})^{\mbox{{\scriptsize inc}}}} =
\frac{{(j^3_{A,s})^{\mbox{{\scriptsize refl}}}} -
{(j^3_{V,s})^{\mbox{{\scriptsize
refl}}}}}{{(j^3_{V,s})^{\mbox{{\scriptsize inc}}}} +
{(j^3_{A,s})^{\mbox{{\scriptsize inc}}}}} \: ,
\\ \nonumber
\\
\label{eq2.56} R_{L \rightarrow R}^{(s)} \!\!& = &\!\! - \frac
{(j^3_{R,s})^{\mbox{{\scriptsize
refl}}}}{(j^3_{L,s})^{\mbox{{\scriptsize inc}}}} =
\frac{{(j^3_{A,s})^{\mbox{{\scriptsize refl}}}} \, + \,
{(j^3_{V,s})^{\mbox{{\scriptsize
refl}}}}}{{(j^3_{A,s})^{\mbox{{\scriptsize inc}}}} \, - \,
{(j^3_{V,s})^{\mbox{{\scriptsize inc}}}}} \: .
\end{eqnarray}
Taking into account the expressions for the currents, we obtain
\begin{equation}
\label{deltaR} \Delta R^{(s)} = 2 R^{(0)}_{s,B} \, ( \,
\delta^{\mbox{{\scriptsize inc}}}_{s,B} \, + \,
\delta^{\mbox{{\scriptsize refl}}}_{s,B} ),
\end{equation}
where we have defined the quantities
\begin{equation}
\label{delta}  \delta^{\mbox{{\scriptsize inc}}}_{s,B} =
\frac{\xi}{2 |\alpha_s|} \left[ \frac {\gamma_-(- \alpha_s,
\beta_s)} {\gamma_+ (\alpha_s,\beta_s)} \, I^{(s)}_2 + \mbox{c.c.}
\right] \! , \;\;\;\; \left. \delta^{\mbox{{\scriptsize
refl}}}_{s,B} = \delta^{\mbox{{\scriptsize inc}}}_{s,B}
\right|_{\beta_s \rightarrow - \beta_s} \! .
\end{equation}
In the case of vanishing magnetic field, $\Delta R^{(s)}$ reduces
to the same expression calculated in Ref.~\cite{funakubo}.

In order to study $\Delta R^{(s)}$ we have to know $f(x)$ and
$g(x)$. As regards $f(x)$ we take the usual bubble wall profile
\begin{equation}
\label{eq2.57} f(x) = \frac{1 + \text{tanh}x}{2} \: .
\end{equation}
With the profile (\ref{eq2.57}), the solutions $\phi_{\pm}^{(\pm
\alpha_s)}(x)$ of the unperturbed equation (\ref{eq2.21}) may be
written as
\begin{eqnarray}
\label{eq2.60} \phi_{\pm}^{(+ \alpha_s)}(y) \!\!& = &\!\! y^{-
\frac{\alpha_s}{2}} (1-y)^{\frac{\beta_s}{2}} \, _2 F_1 \! \left[
\frac{- \alpha_s + \beta_s \mp \xi}{2} + 1 , \frac{-\alpha_s +
\beta_s \pm \xi}{2}, 1 - \alpha_s \, ; y \right] \! ,
\\ \nonumber
\\
\label{eq2.61} \phi_{\pm}^{(- \alpha_s)}(y) \!\!& = &\!\!
y^{\frac{\alpha_s}{2}} (1-y)^{\frac{\beta_s}{2}} \, _2 F_1 \!
\left[ \frac{\alpha_s + \beta_s \mp \xi}{2} + 1 , \frac{\alpha_s +
\beta_s \pm \xi}{2} , 1 + \alpha_s \, ; y \right] \! ,
\end{eqnarray}
where $y = 1 - f(x)$, and $\, _2 F_1(a,b,c;x)$ is the the
hypergeometric function. Now, $\gamma_{\pm}(\alpha_s, \beta_s)$
and $R^{(0)}_{s,B}$ are explicitly given by
\begin{equation}
\label{eq2.28} \gamma_{\pm}(\alpha_s, \beta_s) = \frac{\Gamma(-
\alpha_s + 1) \: \Gamma(- \beta_s)}{\Gamma[(- \alpha_s - \beta_s
\pm \xi)/2] \: \Gamma[(-\alpha_s - \beta_s \mp \xi)/2+1]} \, ,
\end{equation}
and
\begin{equation}
\label{Rzero} R^{(0)}_{s,B} = \frac{\sin [(\pi/2)(\alpha_s -
\beta_s + \xi)] \, \sin [(\pi/2)(\alpha_s -\beta_s - \xi)]}{\sin
[(\pi/2)(\alpha_s + \beta_s + \xi)] \, \sin [(\pi/2)(\alpha_s +
\beta_s - \xi)]} \, .
\end{equation}

The functional form of $g(x)$ is unknown. Following Ref.
\cite{funakubo}, we consider two expressions for $g(x)$:
\begin{eqnarray}
\label{eq2.58} g(x) \!\!& = &\!\! \Delta \theta f^2(x),
\\
\label{eq2.59} g(x) \!\!& = &\!\! \Delta \theta f'(x),
\end{eqnarray}
where the parameter $\Delta \theta$ characterizes the magnitude of
CP-violation.

In Figs.~1-4 we plot $\Delta R^{(s)}/ \Delta \theta$ as a function
of normalized energy $\epsilon^* = \epsilon \, a^* = E/m_0$, at
fixed thickness $a^* = a/m_0 = 1/\xi$ ($a^* = 5$ in Figs.~1 and 3,
$a^* = 1$ in Fig.~2 and 4) for different values of the magnetic
field $b^* = b \, {a^*}^2 = eB/m_0^2$. We take $s=1$ (left panels)
and $s=2$ (right panels), with $g(x) = \Delta \theta f^2(x)$ in
Figs.~1,2, and $g(x) = \Delta \theta f'(x)$ in Figs.~3,4. We found
that $\Delta R^{(s)}$ display the following peculiar properties:

{\it i}) At fixed thickness of the wall (proportional to 1/$a^*$)
and magnetic field, the absolute value of $\Delta R^{(s)}$ goes to
zero when the energy of the incident particles approaches to
infinity;

{\it ii}) The maximum value of $|\Delta R^{(s)}|$ vary both with
the thickness of the wall and with the functional form of $g(x)$.
In particular, we found that at fixed energy and magnetic field,
$|\Delta R^{(s)}|$ is an increasing function of the thickness of
the wall;

{\it iii}) The presence of a magnetic field generates a reflection
asymmetry between spin-up and spin-down particles. In particular,
the effect of the magnetic field is to shift the values of $\Delta
R^{(s)}$ (with respect to the case $B=0$) towards lower energies
in the case $s=1$, and higher energies in the case $s=2$.

The peculiar global quantity in non-local defect mediated
electroweak baryogenesis is the flux of lepton number radiated by
the bubble wall. In the thermal frame of reference, it is given by
(see {\it e.g.} Ref. [16]):
\begin{equation}
\Phi_L = \int \! \frac{d^3 k}{(2 \pi)^3} \; e^{-k/T} L({\textbf
k}) \cos \vartheta_R \, ,
\end{equation}
where $T$ is the temperature and, considering only one type of
particle, $L({\textbf k}) = l |R({\textbf k})|^2$. Here,
$R({\textbf k})$ is the reflection amplitude for a particle of
momentum ${\textbf k}$, $\vartheta_R$ is the angle of reflection
off the advancing wall, $k = |{\textbf k}|$, and $l$ is the lepton
number. In the case of an infinitely planar wall only the motion
of fermions perpendicular to the wall is important, and then we
have $\Phi_L (B) \propto \int \! dE \, e^{-E/T} \Delta R (E,B) \,
$, where $E$ is the energy of the scattered particle, and $\Delta
R = R_{R \rightarrow L} - R_{L \rightarrow R} \,$. Indeed, the
relevant quantity for the generation of a cosmological baryon
asymmetry turns out to be $\Phi_L^{tot}(B) + \Phi_L^{tot}(-B)$,
where
\begin{equation}
\Phi_L^{tot}(B) = \sum_{s=1,2} \Phi_L^{(s)}(B) = \alpha \!
\sum_{s=1,2} \int \! dE \, e^{-E/T} \, \Delta R^{(s)}(E,B) \, .
\end{equation}
Here, $\alpha > 0$ is a numerical factor, and $s$ refers to the
spin of the scattered particle. It is straightforward to verify
that the following transformation law hold:
\begin{equation}
\Delta R^{(s)}(-B) = \Delta R^{(\overline{s})}(B),
\end{equation}
where $\overline{s} = 1$ if $s=2$, and $\overline{s} = 2$ if
$s=1$. Finally, taking into account the previous equations, we
have
\begin{equation}
\Phi_L^{(s)}(B) + \Phi_L^{(s)}(-B) = \Phi_L^{(s)}(B) +
\Phi_L^{(\overline{s})}(B) = \Phi_L^{tot}(B),
\end{equation}
and
\begin{equation}
\Phi_L^{tot}(B) + \Phi_L^{tot}(-B) = 2 \Phi_L^{tot}(B).
\end{equation}
In general, the quantity $\Phi_L^{tot}(B)$ is not null. For
example, from Fig.~1, we get that $\Phi_L^{(1)}(B)$ and
$\Phi_L^{(2)}(B)$ are strictly negative quantities. Hence, in this
case we have $\Phi_L^{tot}(B) = \Phi_L^{(1)}(B) + \Phi_L^{(2)}(B)
< 0$, and then, in general, the total lepton number flux radiated
by an infinitely planar wall is different than zero.
%
%
\renewcommand{\thesection}{\normalsize{\arabic{section}.}}
\section{\normalsize{Conclusions}}
\renewcommand{\thesection}{\arabic{section}}

In this work we have investigate a baryogenesis scheme at a first
order electroweak phase transition in which fermions interact with
a CP-violating thick bubble wall in the background of an uniform
magnetic field perpendicular to the wall.

We have studied the net flux by the difference $\Delta R^{(s)}$ of
the reflection coefficients of the left-handed and right-handed
fermions incident from the unbroken phase at the expanding bubble
wall. We have found that the presence of a magnetic field
generates a reflection asymmetry between spin-up and spin-down
particles. Moreover, we have showed that the only effect of the
magnetic field is to shift the values of $\Delta R^{(s)}$ (with
respect to the case of vanishing magnetic field) towards lower
energies in the case of spin-up particles, and higher energies in
the case of spin-down particles.

In non-local defect mediated electroweak baryogenesis, the total
baryon asymmetry turns out to be proportional to the total lepton
number flux radiated by a bubble wall, $\Phi_L^{tot}(B) -
\Phi_L^{tot}(-B)$. We have showed that, in general, this quantity
is not null. In any case, the cosmological implications of a
non-vanishing lepton number flux are beyond the aim of the present
paper. We deserve such an analysis to a future work.
%
%
\section*{\normalsize{Acknowledgments}}
We would like to thank the referee for his/her insightful comments
about the importance, in non-local defect mediated electroweak
baryogenesis, of the lepton number flux radiated by a bubble wall.
\newpage
%
%
\appendix{\normalsize{\bf {Appendix}}}
\section{\normalsize{Transmitted wave function}}
In this Appendix we calculate the transmitted wave function. In
the same way it is straightforward to calculate the incident and
reflected wave functions.
\\
The wave function to the first order is given by
Eq.~(\ref{eq2.15}), that we rewrite for simplicity:
\begin{equation}
\label{eqA1} \Psi(\tau,x) \simeq \left[ \left(\sigma \epsilon
\gamma^0 + i \gamma^3 \frac{d}{dx} + \xi f + bx \gamma^2 \right)
(\psi^{(0)} + \psi^{(1)}) + i \xi g \gamma_5 \psi^{(0)} \right] \!
e^{-i \sigma \epsilon \tau} .
\end{equation}
Hence, in order to calculate the asymptotic form of
$\Psi(\tau,x)$, it is necessary to know the asymptotic form of
$\psi^{(0)}(x)$ and $\psi^{(1)}(x)$. Taking into account
Eqs.~(\ref{eq2.22}) and (\ref{eq2.29}) we obtain, for positive
energy solutions and in the limit $x \rightarrow +\infty$,
\begin{equation}
\label{eqA3} [\psi^{(0)}(x)]^{+\infty}_{\sigma = +1} = A_s^{(+)}
\, u^s_{+} \, e^{\alpha_s x} .
\end{equation}
As regards $\psi^{(1)}(x)$ we start with Eq.~(\ref{eq2.34}).
Taking into account Eqs.~(\ref{eq2.22}), (\ref{eq2.24}), and
Eqs.~(\ref{eq2.35}), (\ref{eq2.35a}) we have
\begin{equation}
\label{eqA6} [\psi^{(1)}_s(x)]^{+\infty}_{\sigma = +1} = A^{(+)}_s
\, \frac{(-1)^s \xi}{2 \alpha_s} \left[ I_1^{(s)} + c^{(+)}_{-}
I^{(s)}_2 \right] \! u^s_{-} \, e^{\alpha_s x} .
\end{equation}
Remembering that $\lim_{x \rightarrow + \infty} f(x) = 1$, the
transmitted wave function turns out to be
\begin{eqnarray}
\label{eqA9} [\Psi(\tau,x)]^{\mbox{{\scriptsize tran}}}_{\sigma =
+1} \!\!& = &\!\! \left[ \left( \epsilon \gamma^0 + i \gamma^3
\frac{d}{dx} + \xi + bx \gamma^2 \right) \!\! \left( \!\!\!\!\!\!
\phantom{\frac{d}{dx}} \! [\psi^{(0)}_s(x)]^{+ \infty}_{\sigma =
+1} + [\psi^{(1)}_s(x)]^{+ \infty}_{\sigma = +1} \right) \right.
\nonumber
\\
\!\!& + &\!\!  i \xi g_{+} \gamma_5 \, [\psi^{(0)}_s(x)]^{+
\infty}_{\sigma = +1} \left. \!\!\!\!\!\!\!\!\!
\phantom{\frac{d}{dx}} \right] e^{-i \epsilon \tau} .
\end{eqnarray}
Finally, inserting Eqs.~(\ref{eqA3}) and (\ref{eqA6}) into
Eq.~(\ref{eqA9}), we get
\begin{eqnarray}
\label{eqA11} [\Psi_s(\tau,x)]^{\mbox{{\scriptsize tran}}}_{\sigma
= +1} \!\!& = &\!\! A^{(+)}_s \left\{ \left[ \xi - \alpha_s +
\frac{(-1)^s \xi \epsilon}{2 \alpha_s} \, \left( I^{(s)}_1 +
c^{(+)}_{-} I^{(s)}_2 \right) \right] \right. \! u^s_{+} \nonumber
\\
\!\!& + &\!\! \left. \left[ \epsilon + \frac{(-1)^s}{2 \alpha_s}
\, \xi (\xi + \alpha_s) \left( I_1^{(s)} + c^{(+)}_{-} I^{(s)}_2
\right) - (-1)^s \xi \, g_{+} \right] \! u^s_{-} \right\}
\nonumber
\\
\!\!& \times &\!\! e^{-i \epsilon \tau + \alpha_s x} .
\end{eqnarray}
Following the same arguments of Appendix in Ref.~\cite{funakubo},
it is possible to redefine $A^{(+)}_s$ as
\begin{equation}
\label{eqA13} A^{(+)}_s \! \left[ 1 + (-1)^s \, \frac{\xi (\xi +
\alpha_s)}{2 \epsilon \alpha_s} \, c^{(+)}_{-} I^{(s)}_2 \right]
\!\! \left[ 1 - (-1)^s  \frac{\xi}{2 \epsilon} \, g_{-} \right]
\rightarrow A^{(+)}_s .
\end{equation}
Taking into account Eq.~(\ref{eqA13}), it is straightforward to
cast Eq.~(\ref{eqA11}) into the form of Eq.~(\ref{eq2.36}).
\vfill
\newpage
%
%
\renewcommand{\thesection}{\normalsize{\arabic{section}.}}

%
\vfill
\newpage
%
\begin{figure}[H]
\begin{center}
\includegraphics[clip,width=0.49\textwidth]{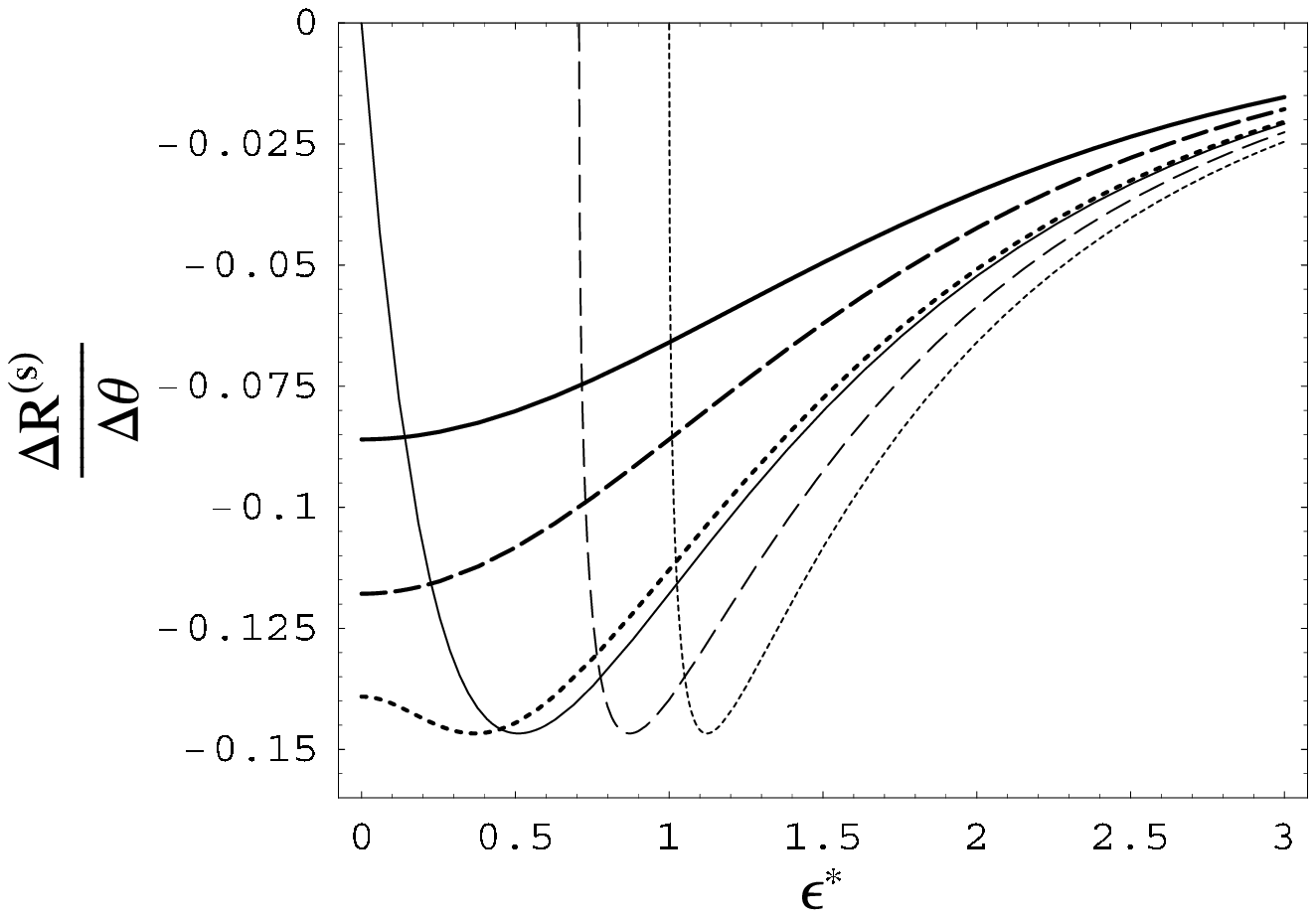}
\includegraphics[clip,width=0.49\textwidth]{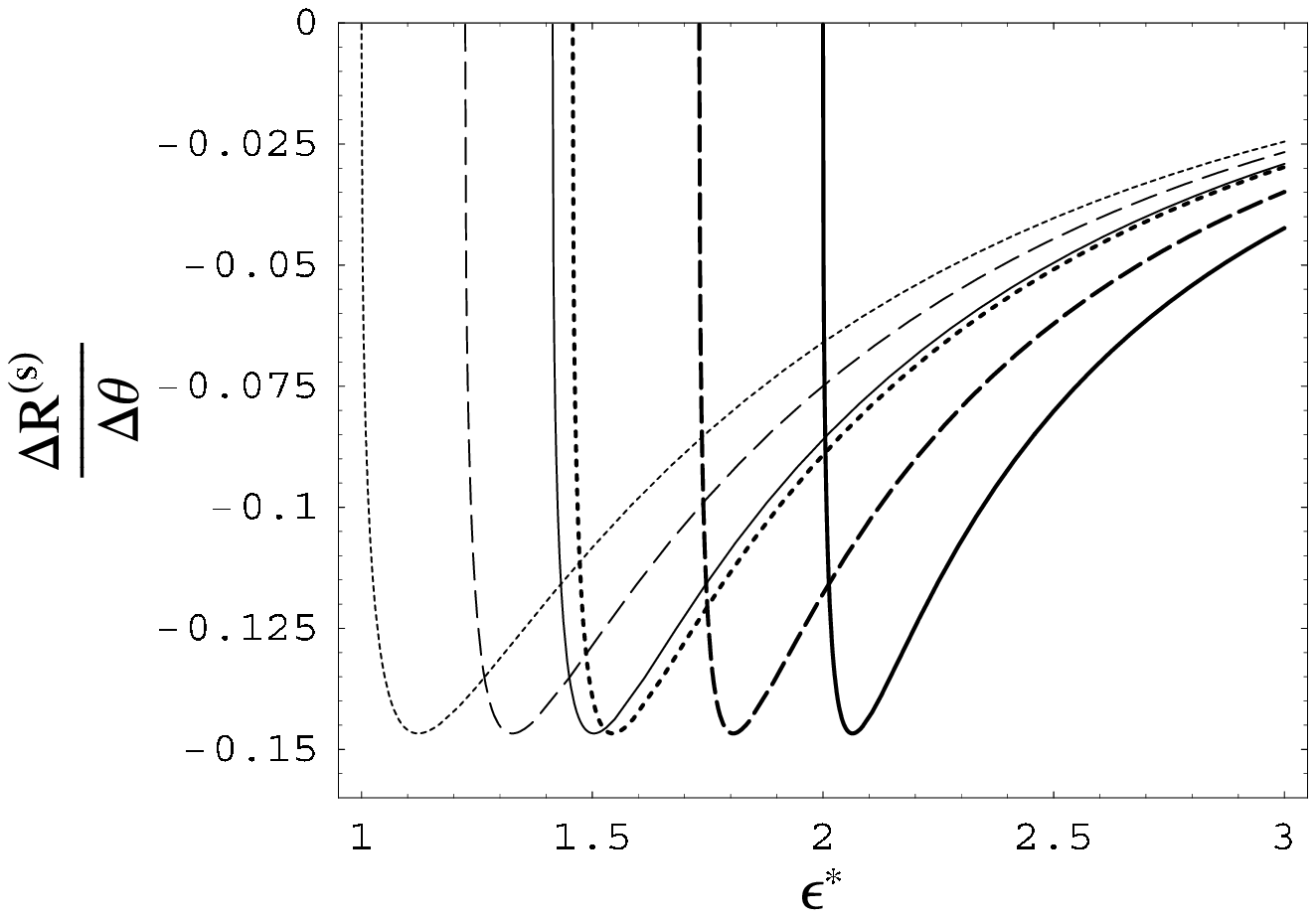}
\caption{$\Delta R^{(s)} / \Delta \theta$ versus $\epsilon^*$,
with $s=1$ (left panel) and $s=2$ (right panel), for different
values of normalized magnetic field $b^*$, in the case $g = \Delta
\theta f^2$ and $a^* = 5$. Thick solid line: $b^* = 3$; thick long
dashed line: $b^* = 2$; thick short dashed line: $b^* = 1.125$;
thin solid line: $b^* = 1$; thin long dashed line: $b^* = 0.5$;
thin short dashed line: $b^* = 0$.} \vskip 1.5truecm
\includegraphics[clip,width=0.49\textwidth]{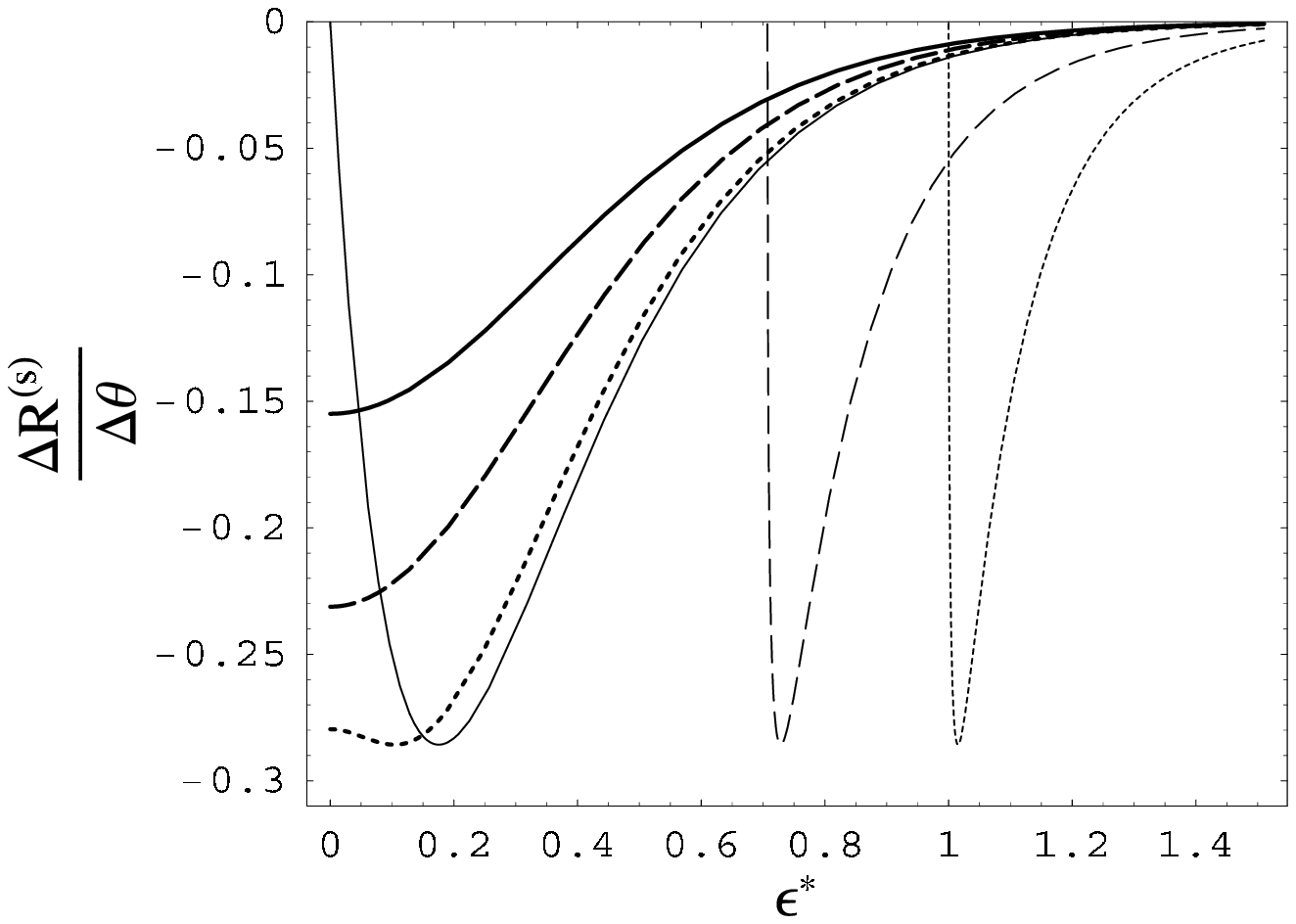}
\includegraphics[clip,width=0.49\textwidth]{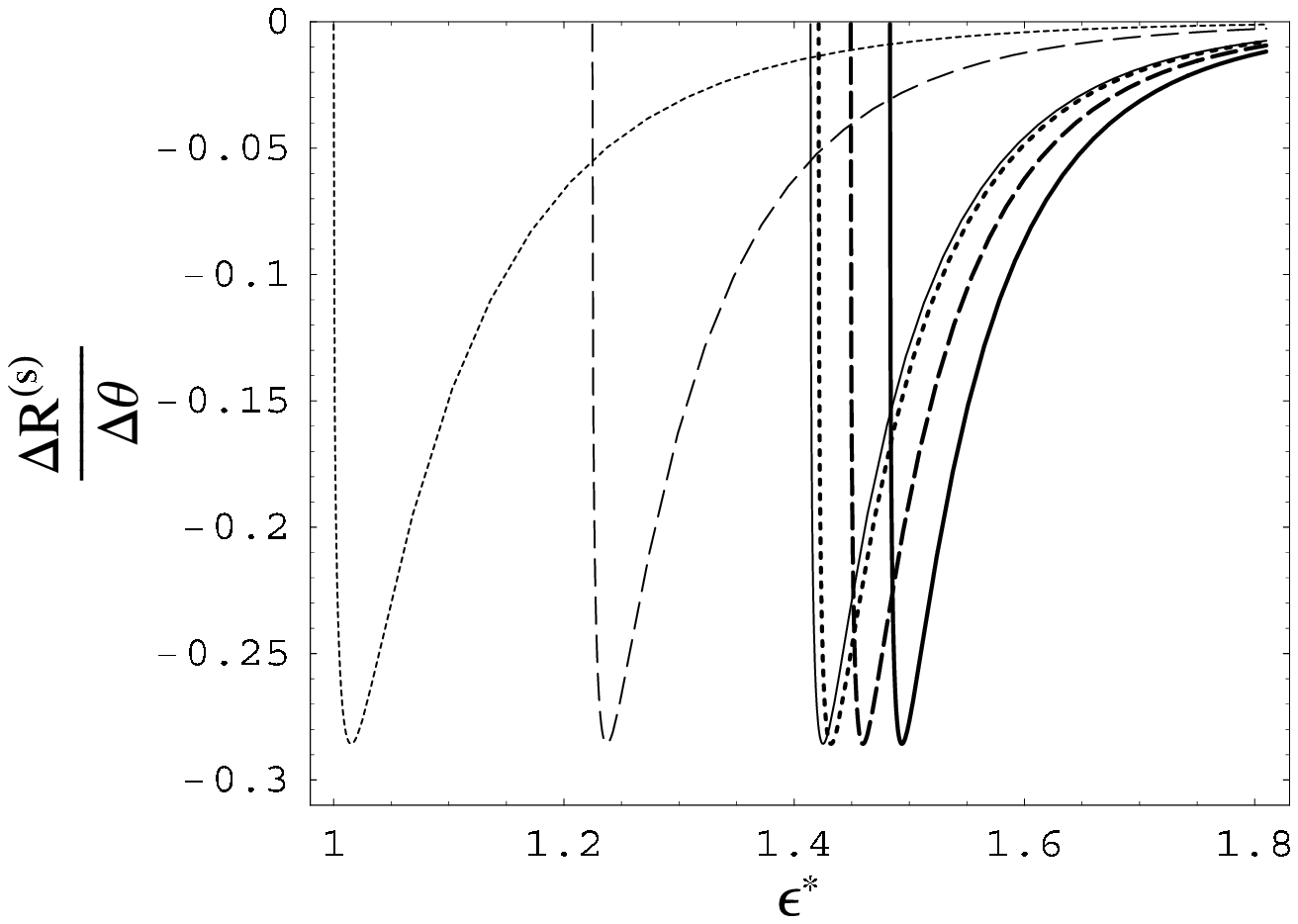}
\caption{$\Delta R^{(s)} / \Delta \theta$ versus $\epsilon^*$,
with $s=1$ (left panel) and $s=2$ (right panel), for different
values of normalized magnetic field $b^*$, in the case $g = \Delta
\theta f^2$ and $a^* = 1$. Thick solid line: $b^* = 1.2$; thick
long dashed line: $b^* = 1.1$; thick short dashed line: $b^* =
1.02$; thin solid line: $b^* = 1$; thin long dashed line: $b^* =
0.5$; thin short dashed line: $b^* = 0$.}
\end{center}
\end{figure}
%
%
\begin{figure}[H]
\begin{center}
\includegraphics[clip,width=0.49\textwidth]{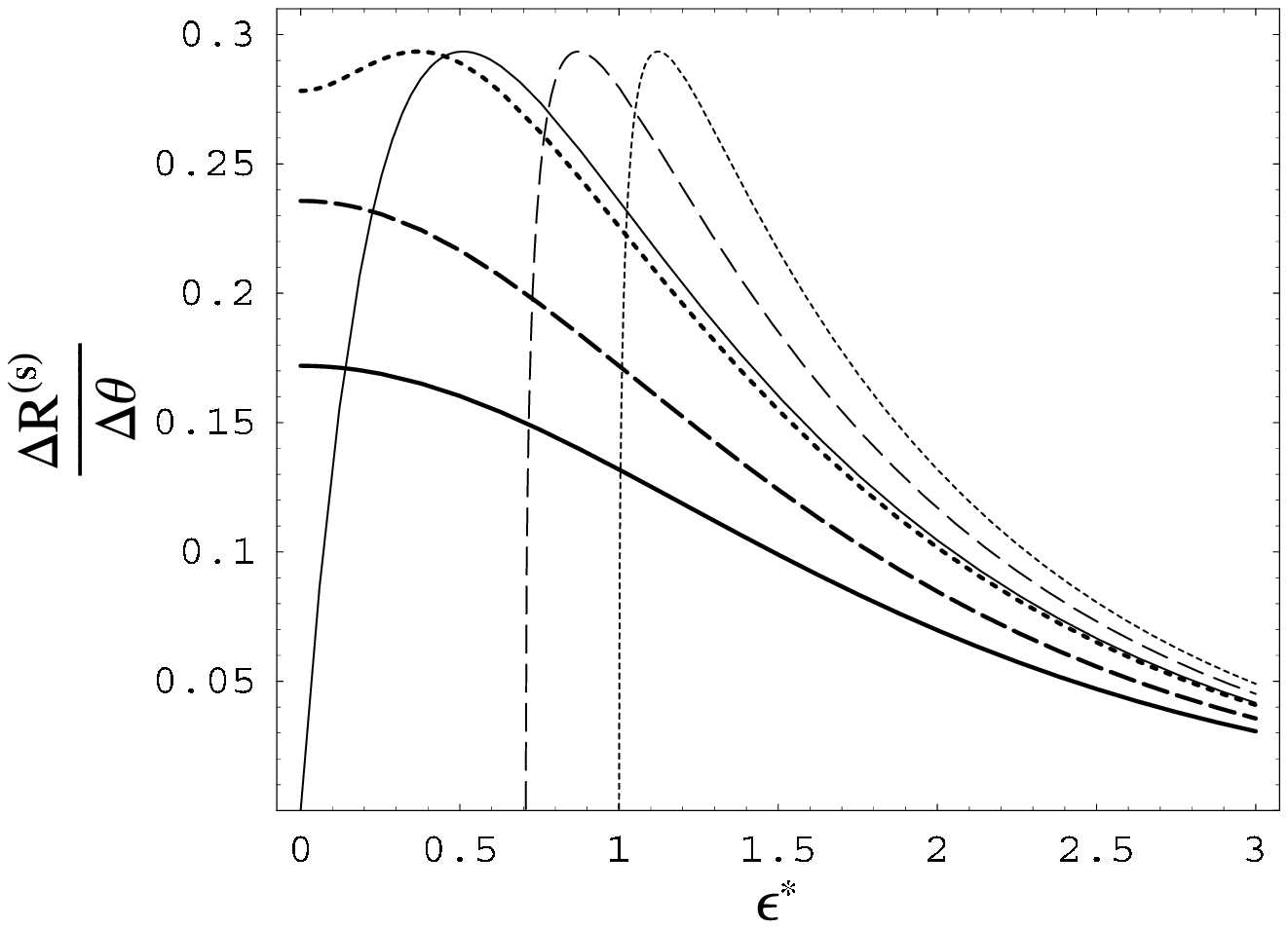}
\includegraphics[clip,width=0.49\textwidth]{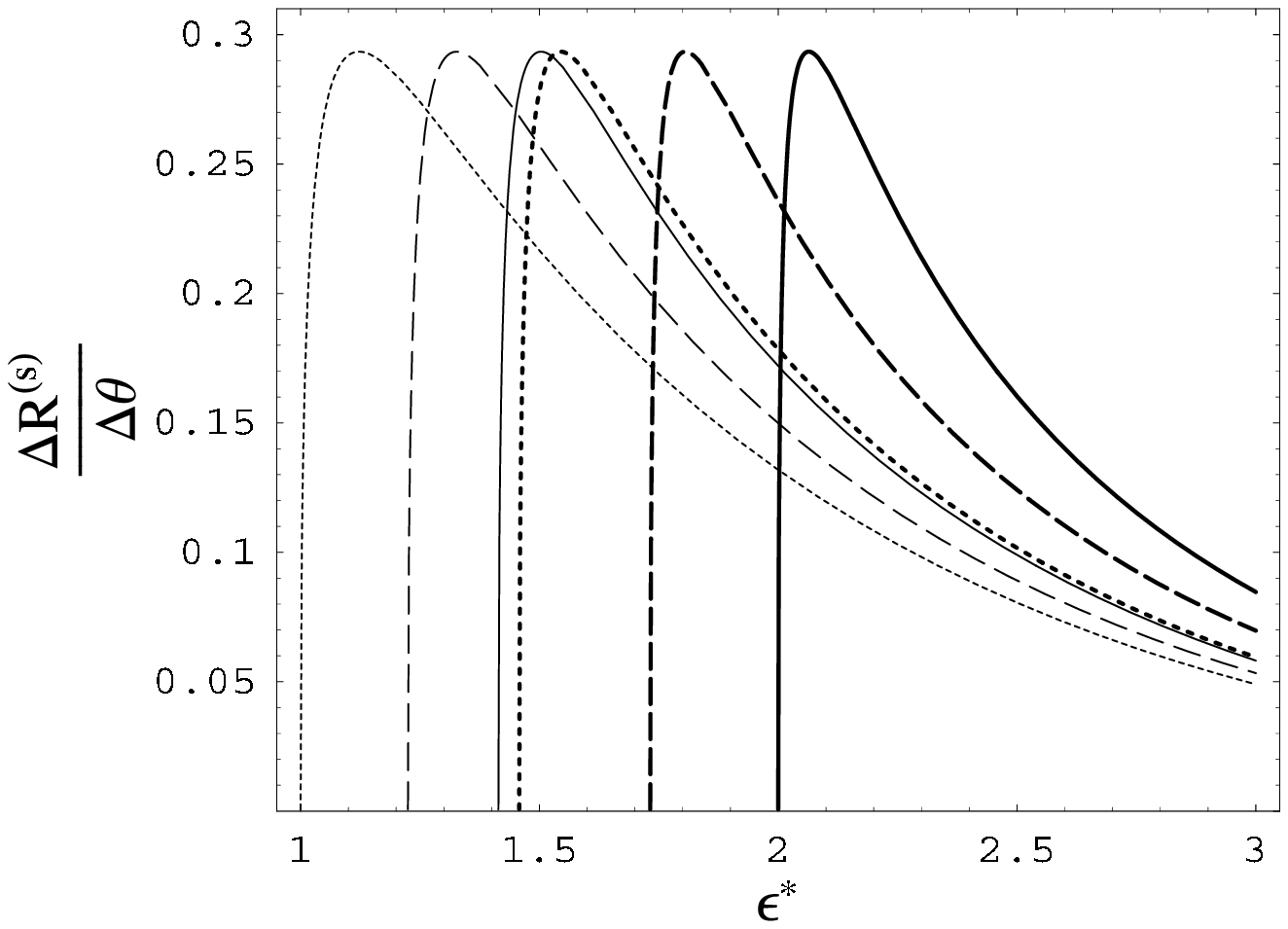}
\caption{$\Delta R^{(s)} / \Delta \theta$ versus $\epsilon^*$,
with $s=1$ (left panel) and $s=2$ (right panel), in the case $g =
\Delta \theta f'$ and $a^* = 5$. The values of $b^*$ are the same
as in Fig.~1.} \vskip 1.5truecm
\includegraphics[clip,width=0.49\textwidth]{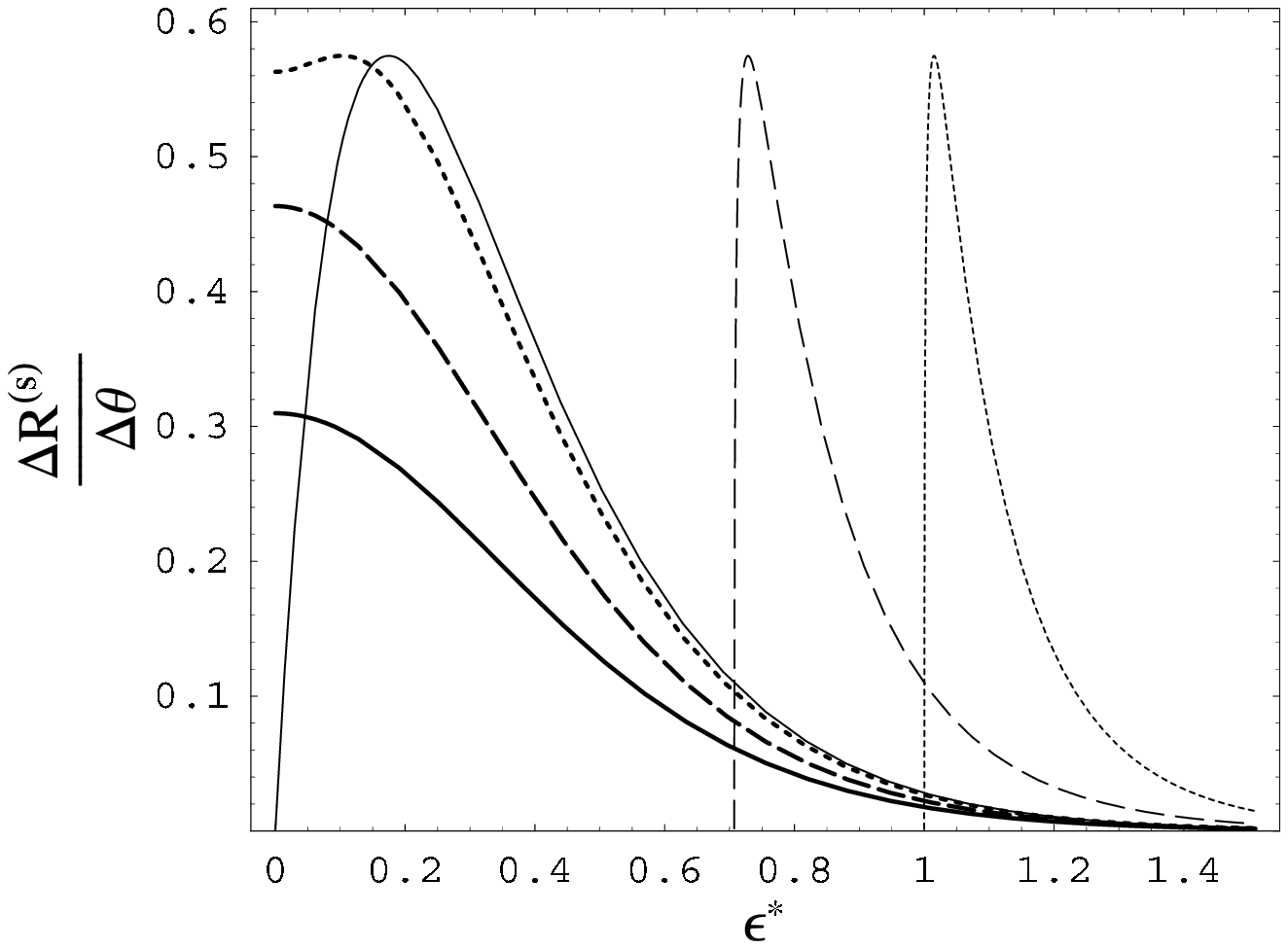}
\includegraphics[clip,width=0.49\textwidth]{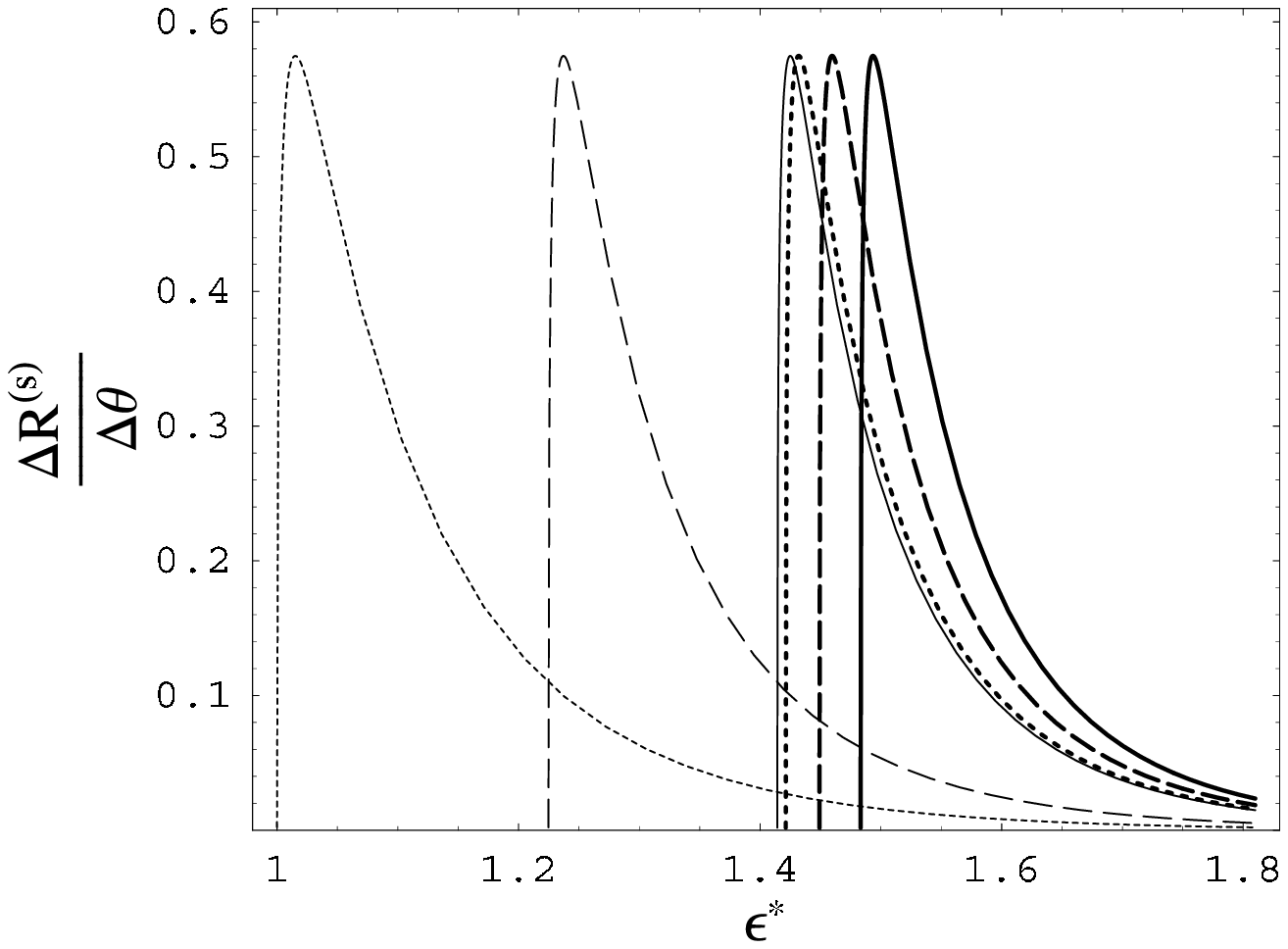}
\caption{$\Delta R^{(s)} / \Delta \theta$ versus $\epsilon^*$,
with $s=1$ (left panel) and $s=2$ (right panel), in the case $g =
\Delta \theta f'$ and $a^* = 1$. The values of $b^*$ are the same
as in Fig.~2.}
\end{center}
\end{figure}
%
%
\end{document}